# Optimal Spectrum Management in Multiuser Interference Channels

Yue Zhao, *Student Member, IEEE,* and Gregory J. Pottie, *Fellow, IEEE*

*Abstract*—In this paper, we study the non-convex problem of continuous frequency optimal spectrum management in multiuser frequency selective interference channels. Firstly, a simple pairwise channel condition for FDMA schemes to achieve all Pareto optimal points of the rate region is derived. It enables fully distributed global optimal decision making on whether any two users should use orthogonal channels. Next, we present in detail an analytical solution to finding the global optimum of sum-rate maximization in two-user symmetric flat channels. Generalizing this solution to frequency selective channels, a convex optimization is established that solves the global optimum. Finally, we show that our method generalizes to *K*-user ($K \geq 2$) weighted sum-rate maximization in asymmetric frequency selective channels, and transform this classic non-convex optimization in the primal domain to an equivalent convex optimization. The complexity is shown to be separable in its dependence on the channel parameters and the power constraints.

*Index Terms*—Optimal spectrum management, multi-user interference channel, FDMA optimality condition, non-convex optimization,

## I. INTRODUCTION

IN multiuser communications systems, interference coupling between different users remains a major problem that limits the system performance. A general multiuser interference channel is depicted in Figure 1, in which each user consists of a transmitter and receiver pair, and there is cross interference coupling between every pair of users. In this paper, we consider the decoding assumption that every receiver *treats the interference from other undesired transmitters as noise*. Some *weak interference* conditions have been found under which treating interference as noise achieves the information theoretic capacity [1] [15] [16]. In general, potentially higher system capacity can be achieved with more complex decoding techniques such as interference cancellation or joint decoding. However, finding the general optimal schemes using these techniques to approach the information theoretic capacity region remains a very hard open problem, although there has been considerable recent progress [10]. Furthermore, these techniques often require that a user knows other users' codebooks, whereas treating interference as noise does not.

We consider the scenario of multiple multicarrier communications systems contending in a common frequency band. (There may sometimes be practical reasons to channelize the resources in some other fashion, e.g. in time. Here, we regard any such alternatives as equivalent to channelizing in the frequency domain [12] [18].) We investigate the optimal *continuous frequency* spectrum and power allocation problem, for which the channel frequency responses and the users' power spectral density (PSD) can be any *piecewise bounded continuous functions* of the frequency over a finite band. The continuous frequency problem is an infinite dimensional optimization, except that in the special case of a *frequency flat* channel response, it has been shown to have a finite number of dimensions [9] [17]. Despite the infinite number of dimensions, significant insights can still be provided from solving the optimal solution in this continuous frequency form, as shown in the later sections of this paper.

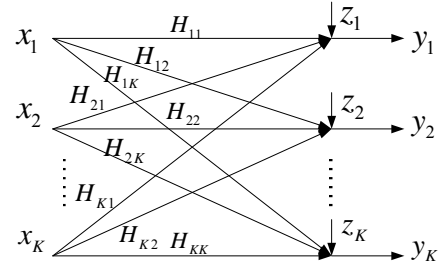

Fig. 1. Multiuser Interference Channel

In practical systems, a discrete frequency model with a finite number of sub-carriers is often assumed, and the power allocations within each sub-carrier are required to be flat. For a variety of objective functions, the non-convex optimization of spectrum and power with the discrete frequency model has been shown to be NP complete in the number of users even for the single carrier case [14]. For the single carrier *sum-rate* maximization problem, two special cases have been solved: the two-user case of all channel parameters [8] [11], and the *K*-user ($K \geq 2$) case of fully symmetric channels [2]. For the multicarrier weighted sum-rate maximization problem, generally known as *spectrum management* or *spectrum balancing*, there has been a great amount of research addressing the non-convexity and NP completeness. With sufficient primal objective relaxations, the problem can be approximated as convex optimizations [6] [7]. For solving the original non-convex optimization, dual decomposition methods have been widely applied to decompose the problem in frequency [4] [5] [21] [24]. While these methods effectively reduced the scale of the problem to solve, two remaining issues are as follows. *1)* While the dual master problem is a convex optimization (which can be solved by e.g. subgradient method [21],) the single carrier *sub-problem* is still an NP-complete non-convex optimization. *2)* The dual optimal solution does not necessarily give a primal optimal solution. Addressing the second issue, a significant result is that the duality gap of the spectrum management problem goes to zero as the number of sub-channels goes to infinity, under mild technical conditions [21][14]. Our results will be connected to this result at the end of this paper.

There are essentially two strategies for multiple users to co-exist: *FDMA and frequency sharing (overlapping).* As the cross coupling varies from being extremely strong to extremely weak, the preferable co-existence strategies intuitively shift from complete avoidance (FDMA) to pure frequency sharing. We start from the strong coupling scenario, and investigate the weakest interference condition under which FDMA is still guaranteed to be optimal, regardless of the power constraints. In the literature, a relatively strong *pairwise* coupling condition for FDMA to achieve all Pareto optimal points of the rate region is derived with the continuous frequency model [9]. By pairwise we mean that whether two users should be orthogonalized in frequency only depends on the interference condition between those two users. For *sum-rate* maximization, the required coupling strengths for FDMA to be optimal are further lowered, approaching roughly the weakest possible [13]. However, this condition is derived in a *group-wise* form, requiring the couplings between all users to be sufficiently strong.

In this paper, by analyzing the continuous frequency model, the weakest possible *pairwise* condition for FDMA to achieve *all* Pareto optimal points of the rate region is proved: for any two (among all of the $K$) users, as long as the two normalized cross channel gains between them are both larger than or equal to 1/2, an FDMA allocation between these two users benefits *every one of the $K$ users.* When the cross channel gain is less than 1/2 in symmetric channels, we precisely characterize the non-empty power constraint region within which frequency sharing between two users leads to a higher rate than an FDMA allocation between them.

For the general non-convex optimization of spectrum management, we develop a new method that transforms the problem in the *primal* domain into an *equivalent convex optimization.*

We begin with *sum-rate* maximization in two-user symmetric flat channels. We show that the optimal spectrum management can be solved by computing a convex hull function (which boils down to solving a closed form equation.) As a result, the optimal spectrum management always consists of one sub-band of flat frequency sharing and one sub-band of flat FDMA. This sets up our more general results. The optimal solution for the *sum-rate* maximization was also independently derived in [17] for *two-user asymmetric* flat channels, and in [2] for *K-user* ($K \geq 2$) *symmetric* flat channels.

We first generalize our results to two-user symmetric *frequency selective* channels, and show that a convex relaxation of the original non-concave objective actually leads to the *same* optimal value as the original problem.

Next, we generalize our results to *K-user asymmetric* flat channels for arbitrary *weighted sum-rate* maximization, and show that the optimal solution can be found by computing a convex hull function.

Finally, we combine the ideas of these generalizations, and establish the equivalent primal domain convex optimization for the spectrum management problem in its general form, i.e., *arbitrary weighted sum-rate maximization for K-user* ($K \geq 2$) *asymmetric frequency selective channels.*

The rest of the paper is organized as follows. The problem model is established in Section II. In Section III, we discuss the channel conditions under which FDMA schemes can achieve all Pareto optimal rate tuple. In Section IV, we solve the sum-rate maximization in two-user symmetric (potentially frequency selective) channels. In Section V, we extend our method to the most general cases, and show that the continuous frequency optimal spectrum management scheme can be solved by a primal domain convex optimization. Conclusions are drawn in Section VI.

## II. CHANNEL MODEL AND TWO BASIC CO-EXISTENCE STRATEGIES

### A. Interference Channel Model and the Rate Density Function

As depicted in Figure 1, a $K$-user Gaussian interference channel is modeled by

$$y_i = H_{ii}x_i + \sum_{j \neq i} x_j H_{ji} + z_i, \quad i = 1, 2, ..., K .$$

where $x_i$ is the transmitted signal of user $i$, and $y_i$ is the received signal of user $i$ including additive Gaussian noise $z_i$, (a user corresponds to a transmitter and receiver pair). $H_{ii}$ is the direct channel gain from the transmitter to the receiver of user $i$. $H_{ji}$ is the cross channel gain from the transmitter of user $j$ to the receiver of user $i$. For the purposes of the analysis in this paper, without loss of generality (WLOG), we assume that the channel is over a unit bandwidth frequency band [0, 1]. The results derived directly generalize to frequency bands with arbitrary bandwidths.

The frequency selective $H_{ii}$ and $H_{ji}$ are denoted by $H_{ii}(f)$ and $H_{ji}(f), f \in [0,1]$. Denote the transmit PSD of user $i$ by $p_i(f)$, and the noise PSD at receiver $i$ by $\sigma_i(f)$. We assume that $H_{ji}(f), \sigma_i(f), p_i(f), \forall i, j$ are all *piecewise bounded continuous* functions over the band $f \in [0,1]$. Furthermore, we assume that all functions appearing in this paper have *a finite number of discontinuities.*

We assume that every user uses a random Gaussian codebook, and only decodes the signal from its own transmitter, treating interference from other transmitters as noise. Employing the Shannon capacity formula for Gaussian channels, we have the following achievable rate for user $i \, (=1,2,...,K)$ :

$$R_i = \int_0^1 \log\left(1 + \frac{p_i(f)\left|H_{ii}(f)\right|^2}{\sigma_i(f) + \sum_{j\neq i} p_j(f)\left|H_{ji}(f)\right|^2}\right) df$$

$$= \int_0^1 \log\left(1 + \frac{p_i(f)}{n_i(f) + \sum_{j\neq i} p_j(f)\alpha_{ji}(f)}\right) df,$$

where $\alpha_{ji}(f) \triangleq \dfrac{\left|H_{ji}(f)\right|^2}{\left|H_{ii}(f)\right|^2}$, $n_i(f) \triangleq \dfrac{\sigma_i(f)}{\left|H_{ii}(f)\right|^2}$ are the cross channel gains and the noise power normalized by the direct channel gains. We further make a technical assumption that

$$\exists n_\varepsilon > 0, \ \text{ s.t. } \ \forall f \in [0,1], \ \ n_i(f) \geq n_\varepsilon, \forall i = 1,2,...,K. \quad (1)$$

which naturally holds in all physical channels.

To reach any Pareto optimal point of the $K$-user rate region, we optimize the power and spectrum allocation functions (also termed as the spectrum management schemes):

$$\mathbf{p}(f) \triangleq [p_1(f), p_2(f),..., p_K(f)]^T, \forall f \in [0,1].$$

As we consider the continuous frequency model, define

*Definition 1:* $\forall f \in [0,1]$, with $\mathbf{P} = [P_1, P_2,..., P_K]^T$,

$$r_i(\mathbf{P}, f) \triangleq \log\left(1 + \frac{P_i}{n_i(f) + \sum_{j\neq i} P_j \alpha_{ji}(f)}\right).$$

Now, we have the *rate density function* of user $i$ at frequency $f$ :

$$r_i(\mathbf{p}(f), f) = \log\left(1 + \frac{p_i(f)}{n_i(f) + \sum_{j\neq i} p_j(f)\alpha_{ji}(f)}\right),$$

and $\mathbf{r}(\mathbf{p}(f), f) \triangleq [r_1(\mathbf{p}(f), f), r_2(\mathbf{p}(f), f),..., r_K(\mathbf{p}(f), f)]^T$.

Accordingly, $R_i = \int_0^1 r_i(\mathbf{p}(f), f)\, df, i = 1,2,...,K.$

### B. Piecewise Continuous Functions as Limits of Piecewise Flat Functions

We consider the channel responses and power allocations as piecewise bounded continuous functions of frequency. Intuitively, one may approximate continuous functions by piecewise constant functions, by subdividing the support (frequency) to a sufficiently large number of small pieces. We make use of this idea in later sections, and provide a technical lemma for this purpose.

*Lemma 1 (Approximation Lemma):*

Given $\mathbf{p}(f), \{\alpha_{ji}(f)\}, \{n_i(f)\}, f \in [0,1]$, all piecewise bounded continuous. For any utility function $U(\mathbf{p}, \boldsymbol{\alpha}, \mathbf{n})$ that is a uniformly continuous function of $\{\{p_i\}, \{\alpha_{ji}\}, \{n_i\}\}$,

$\forall \varepsilon > 0$, there exists a set of *piecewise flat* power allocation functions and channel responses,

$$\overline{\mathbf{p}}(f) = [\overline{p}_1(f),...,\overline{p}_K(f)]^T, \{\overline{\alpha}_{ji}(f)\}, \{\overline{n}_i(f)\}, f \in [0,1],$$

for which the band is divided into $M \, (<\infty)$ intervals $I_1,...,I_M$, $I_m = [f_{m-1}, f_m]$, with $f_0 = 0, f_M = 1, f_{m-1} < f_m$, and $\forall m = 1,2,...,M$,

$$\begin{cases} \overline{\mathbf{p}}(f) = \mathbf{P}(m), \ \forall f \in I_m, \\ \overline{\alpha}_{ji}(f) = \alpha_{ji}(m), \ \overline{n}_i(f) = n_i(m), \forall f \in I_m, \forall i, j. \end{cases}$$

where $\mathbf{P}(m) = [P_1(m),..., P_K(m)]^T, \{\alpha_{ji}(m)\}, \{n_i(m)\}$ are *constants* that only depend on the interval index $m$,
*such that the following three properties hold:*

P1. $\forall f \in [0,1], \ \overline{p}_i(f) \leq p_i(f), \forall i = 1,2,...,K,$

P2. $\forall f \in [0,1], \ \overline{\alpha}_{ji}(f) \geq \alpha_{ji}(f), \forall i \neq j, \ \overline{n}_i(f) \geq n_i(f), \forall i,$

P3. $\forall f \in [0,1], \ \left|U(\overline{\mathbf{p}}(f), \overline{\boldsymbol{\alpha}}(f), \overline{\mathbf{n}}(f)) - U(\mathbf{p}(f), \boldsymbol{\alpha}(f), \mathbf{n}(f))\right| < \varepsilon.$

*Proof:* Details are relegated to Appendix A. ∎

From now on, we name the $\overline{\mathbf{p}}(f), \{\overline{\alpha}_{ji}(f)\}$ and $\{\overline{n}_i(f)\}$ found in Lemma 1 a "*piecewise flat $\varepsilon$ - approximation*".

*Remark 1:* Property P1 ensures that the approximate piecewise flat *power* allocations consume *less* power than the original ones. Property P2 ensures that the approximate piecewise flat *channel* responses are "*worse*" than the original ones (as the cross channel gains and the noise power are all stronger, and interference is treated as noise.) Nonetheless, property P3 ensures that under these "adverse" conditions, these approximations can still achieve the original utility $U$ arbitrarily closely.

With finite power constraints and non-degenerate channel parameters (1), most utility functions considered in practice (e.g. a weighted sum-rate) satisfy the uniform continuity condition of $U(\mathbf{p}, \boldsymbol{\alpha}, \mathbf{n})$.

### C. Two Basic Co-existence Strategies and One Basic Transformation

There are essentially two co-existence strategies for users to reside in a common band: frequency sharing and FDMA. We introduce two basic forms of these two strategies: *Flat Frequency Sharing* and *Flat FDMA*, both defined in flat channels. We will see that these two basic strategies are the building blocks of general non-flat co-existence strategies in frequency selective channels.

Consider a two-user flat channel: $\forall f \in [0, 1]$,

$$n_1(f) = n_1, \ n_2(f) = n_2, \ \alpha_{21}(f) = \alpha_{21}, \ \alpha_{12}(f) = \alpha_{12}. \quad (2)$$

*Definition 2:* A flat frequency sharing scheme of two users is any power allocation in the form of

$$p_1(f) = p_1, \ p_2(f) = p_2, \ \forall f \in [0,1]. \quad (3)$$

*Definition 3:* A flat FDMA scheme of two users is any power allocation in the form of

$$\begin{cases} p_1(f)p_2(f) = 0 \\ p_1(f) + p_2(f) = p \end{cases}, \ \forall f \in [0,1]$$

*Definition 4:* The *flat FDMA reallocation* is the following

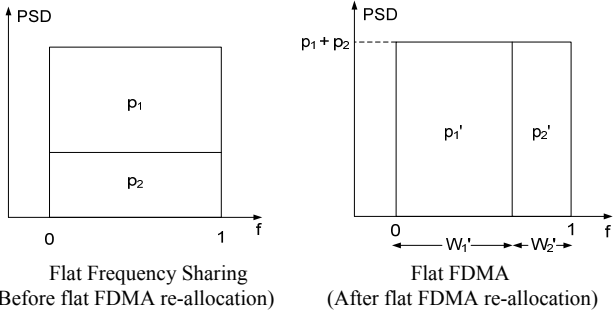

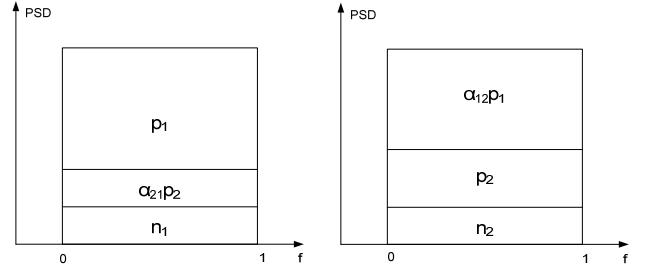

Fig. 2. Power allocations of flat frequency sharing and flat FDMA, also an illustration of flat FDMA re-allocation.

Fig. 3. The PSD composition at receiver 1 and receiver 2

*power invariant* transform that reallocates the power of the two users from a flat frequency sharing scheme to a flat FDMA:

User 1 re-allocates all of its power within a sub-band $W_1' = p_1 / (p_1 + p_2)$ with a flat PSD $p_1' = p_1 + p_2$; user 2 re-allocates all of its power within another disjoint sub-band $W_2' = p_2 / (p_1 + p_2)$ with the same flat PSD $p_2' = p_1 + p_2$.

Illustrations of the power allocations of the two basic co-existence strategies before and after a flat FDMA re-allocation are depicted in Figure 2. Clearly, the total power of each user does not change after this re-allocation.

Similarly, flat frequency sharing schemes, flat FDMA schemes, and flat FDMA re-allocation can be defined for any $k$ ($k = 1, 2, 3, ...$) users. ($k = 1$ is the degraded case in which flat frequency sharing is the same as flat FDMA.)

*Remark 2:* A flat FDMA scheme is mathematically the same as multiple disjoint bands each seeing a flat frequency sharing of only *one* user. Thus, it is actually sufficient to only define flat frequency sharing schemes of any $k$ ($k = 1, 2, 3, ...$) users, *without* introducing the definition of flat FDMA schemes. This alternative approach is used later in Section V for the general optimization in $K$-user frequency selective channels. Here, flat FDMA and flat FDMA re-allocation are explicitly defined, because they offer clear intuitions for optimizing spectrum management as will be shown in Sections III and IV.

## III. THE CONDITIONS FOR THE OPTIMALITY OF FDMA

In this section, we investigate the conditions under which the optimal spectrum and power allocation is FDMA. We show that our results apply to all Pareto optimal points of the achievable rate region.

Firstly, we show a coupling condition under which FDMA achieves all Pareto optimal rate tuples within a group of strongly coupled users. In real communication networks, however, there are usually users not strongly enough coupled with some other users. For these users outside the strongly coupled group, we show that they always benefit from an FDMA allocation within the strongly coupled group. These results lead to the following simple *pairwise* condition: for any two of the $K$ users, as long as the normalized cross channel gains between them are both larger than or equal to 1/2, *every one* of the $K$ users will benefit from an FDMA allocation between these two users.

### A. The Optimality of FDMA within Strongly Coupled Users

In this section, we prove a sufficient condition for $K$-user interference channels under which FDMA among all users can achieve any Pareto optimal rate tuple. This condition requires that between every pair of users, the normalized cross channel gains must be stronger than 1/2. We begin with two-user flat channels, and extend the results to $K$-user frequency selective channels.

*Theorem 1:* Consider a two-user flat interference channel (2). Suppose the two users co-exist in a flat frequency sharing manner (3). If $\alpha_{12} \geq 1/2$ *and* $\alpha_{21} \geq 1/2$, then with a flat FDMA power re-allocation, *both* users' rates will be higher (or unchanged.)

Before proving Theorem 1, we provide the following lemma:

*Lemma 2:* Let $f(x) = \frac{1}{x} \log(\frac{c+x}{c-x})$, $c > 1$, then

$$f(1) \geq f(x), \quad \forall x \in (0, 1].$$

*Proof:* See Appendix B. ∎

*Proof of Theorem 1:*

The received PSD of the desired signal, interference and noise at both receivers are depicted in Figure 3. The rates of user 1 and user 2 are

$$R_1 = \log\left(1 + \frac{p_1}{n_1 + p_2 \alpha_{21}}\right), R_2 = \log\left(1 + \frac{p_2}{n_2 + p_1 \alpha_{12}}\right).$$

With a flat FDMA power re-allocation (c.f. Figure 2),

$$W_1' = \frac{p_1}{p_1 + p_2}, \ p_1' = p_1 + p_2, \ W_2' = \frac{p_2}{p_1 + p_2}, \ p_2' = p_1 + p_2.$$

Denote user 1's rate after re-allocation by

$$R_1' = W_1' \log\left(1 + \frac{p_1'}{n_1}\right) = \frac{p_1}{p_1 + p_2} \log\left(1 + \frac{p_1 + p_2}{n_1}\right).$$

Notice that $R_1 = \log\left(1 + \frac{\hat{p}_1}{\hat{n}_1 + \hat{p}_2 \alpha_{21}}\right)$, $R_1' = \hat{p}_1 \log\left(1 + \frac{1}{\hat{n}_1}\right)$,

where $\hat{p}_1 \triangleq \frac{p_1}{p_1 + p_2}$, $\hat{p}_2 \triangleq \frac{p_2}{p_1 + p_2}$, $\hat{n}_1 \triangleq \frac{n_1}{p_1 + p_2}$ are the power and noise normalized by the *sum-power*. Note that $\hat{p}_1 + \hat{p}_2 = 1$.

We want to show that if $\alpha_{21} \geq 1/2$, we have $R_1' \geq R_1$.

Since $R_1|_{\alpha_{21} = \frac{1}{2}} \geq R_1|_{\alpha_{21} > \frac{1}{2}}$, it is sufficient to show that

$$\alpha_{21} = 1/2 \Rightarrow R_1' \geq R_1 .$$

With $\alpha_{21} = 1/2$, $\hat{p}_1 + \hat{p}_2 = 1$, it is straightforward to show that

$$R_1' \geq R_1 \Leftrightarrow \log\left(\frac{n_1+1}{n_1}\right) \geq \frac{1}{\hat{p}_1}\log\left(\frac{2n_1+1+\hat{p}_1}{2n_1+1-\hat{p}_1}\right).$$

Notice that $\log\left(\dfrac{n_1+1}{n_1}\right) = \dfrac{1}{1}\log\left(\dfrac{2n_1+1+1}{2n_1+1-1}\right)$. Define function $f(x) = \dfrac{1}{x}\log\left(\dfrac{c+x}{c-x}\right)$, where $c = 2n_1+1$. Thus,

$$R_2' \geq R_1 \Leftrightarrow f(1) \geq f(\hat{p}_1) .$$

Since the normalized power $\hat{p}_1 \in (0,1]$, from Lemma 2, $f(1) \geq f(\hat{p}_1)$. Thus, we conclude that $\alpha_{21} \geq \dfrac{1}{2} \Rightarrow R_1' \geq R_1$.

Similarly, we also have $\alpha_{12} \geq 1/2 \Rightarrow R_2' \geq R_2$.

Therefore, *for both users*, a flat FDMA power re-allocation leads to rates higher than or equal to a flat frequency sharing, when $\alpha_{12} \geq 1/2$ and $\alpha_{21} \geq 1/2$. ∎

Theorem 1 can be generalized to the $K$-user case as follows.

*Corollary 1.1:* Consider a $K$-user flat interference channel, $n_i(f) = n_i$, $\alpha_{ji}(f) = \alpha_{ji}$. Suppose the $K$ users co-exist in a flat frequency sharing manner: $\mathbf{p}_i(f) = \mathbf{p}_i, \forall f \in [0,1]$. If $\alpha_{ji} \geq 1/2$, $\forall j \neq i$, then with a flat FDMA power re-allocation, *all* users' rates will be higher or unchanged.

*Proof:* See Appendix B. ∎

This sufficient condition can also be immediately generalized to frequency selective channels.

*Corollary 1.2:* Consider a $K$-user frequency selective interference channel. Suppose we have an arbitrary spectrum and power allocation scheme $\mathbf{p}(f)$ with some frequency sharing (overlapping) in the band. If $\alpha_{ji}(f) \geq 1/2$, $\forall j \neq i$, $\forall f \in [0,1]$, we can always find an *FDMA* power re-allocation scheme $\tilde{\mathbf{p}}(f)$, satisfying $\int_0^1 \tilde{p}_i(f)df = \int_0^1 p_i(f)df$, $i = 1,...,K$, with which *all* user's rates are higher or unchanged.

*Proof:* The proof is immediate as the strong coupling condition is for *all* frequency: $\forall f \in [0,1], \alpha_{ji}(f) \geq 1/2$, $\forall j \neq i$. ∎

We conclude this sub-section as follows: Pick any sub-band $[f_1, f_2]$, as long as all the users are strongly coupled by $\alpha_{ji}(f) \geq 1/2$, $\forall j \neq i$, $\forall f \in [f_1, f_2]$, then for any power allocation scheme having frequency sharing used anywhere within $[f_1, f_2]$, there always exists an FDMA power re-allocation scheme (with every user's total power unchanged) that leads to a rate higher than or equal to the original sharing scheme *for every user*.

### B. FDMA Within a Subset of Users Benefits All Other Users

We have seen that by properly separating a group of strongly coupled users to orthogonal channels, every user among them will have a rate higher than or equal to the rate of any frequency sharing (overlapping) scheme. In this section, we show that an FDMA allocation among a group of users *also benefits every other user outside this group*. This result completes the fundamental fact that, to achieve *any $K$-user Pareto optimal rate tuple*, all the strongly coupled users (among all the users) must be separated into disjoint frequency bands.

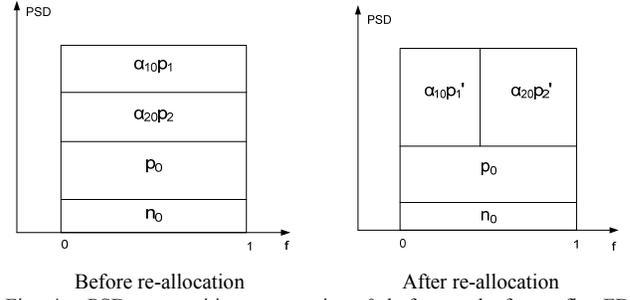

Fig. 4. PSD compositions at receiver 0 before and after a flat FDMA re-allocation of user 1 and user 2.

We begin with the two-interferer flat channels, and extend the results to $K$-interferer frequency selective channels.

*Lemma 3:* Consider a three-user (one user + two interferers) flat channel: $n_i(f) = n_i$, $\alpha_{ji}(f) = \alpha_{ji}$. Suppose the three users co-exist in a flat frequency sharing manner: $p_i(f) = p_i$, $\forall f \in [0,1]$, $i = 0,1,2$. From user 0's perspective, a flat FDMA power re-allocation of its two interferers, namely user 1 and user 2, always leads to a rate higher than or equal to the original rate *for user 0*.

*Proof:* At the receiver of user 0, the received PSDs before and after the flat FDMA power re-allocation of its interferers are depicted in Figure 4. User 0's rates before and after the re-allocation are

$$R_0 = \log\left(1 + \frac{p_0}{\alpha_{10}p_1 + \alpha_{20}p_2 + n_0}\right),$$

$$R_0' = \frac{p_1}{p_1+p_2}\left(1 + \frac{p_0}{\alpha_{10}(p_1+p_2)+n_0}\right)$$
$$+ \frac{p_2}{p_1+p_2}\left(1 + \frac{p_0}{\alpha_{20}(p_1+p_2)+n_0}\right).$$

With straightforward calculations, one can verify that the function $\log(1+\dfrac{P}{I+N})$ is convex in $I$. Therefore, By Jensen's Inequality, $R_0' \geq R_0$, $\forall p_1, p_2 \geq 0$. ∎

Lemma 3 can be generalized to an arbitrary number of users as in the following corollary.

*Corollary 2.1:* Consider a $K+1$-user (one user + $K$ interferers) flat channel: $n_i(f) = n_i$, $\alpha_{ji}(f) = \alpha_{ji}$. Suppose the $K+1$ users co-exist in a flat frequency sharing manner: $\mathbf{p}(f) = \mathbf{p}$, $\forall f \in [0,1]$. From user 0's perspective, a flat FDMA power re-allocation of its $K$ interferers, namely user 1, user 2, $\dots$, user $K$, always leads to a rate higher than or equal to the original rate for user 0.

*Proof:* Similarly to the proof of Lemma 3, it follows from the convexity of $\log(1+\dfrac{P}{I+N})$ in $I$. ∎

Finally, the benefits of an FDMA within a subset of users to the other users can be generalized to frequency selective channels.

*Corollary 2.2:* Consider an $K+1$-user (one user + $K$ interferers) frequency selective channel. Suppose we have an arbitrary spectrum and power allocation scheme $p_i(f)$, $i=0,1,2,...,K$, in which user 1, ..., user $K$ are *not* completely FDMA. Then, from user 0's perspective, there is always a corresponding FDMA power re-allocation of its $K$ interferers, namely user 1, ..., user $K$, that leads to a rate higher than or equal to the original rate for user 0.

*Proof:* $\forall \varepsilon > 0$, by Lemma 1, take a piecewise flat $\varepsilon$-approximation $\overline{\mathbf{p}}(f)$, $\left\{\overline{\alpha}_{ji}(f)\right\}$ and $\left\{\overline{n}_i(f)\right\}$, s.t.

$$\left|\overline{R}_0 - R_0\right| < \varepsilon \,,$$

where $\overline{R}_0$ is user 0's rate computed with $\overline{\mathbf{p}}(f)$, $\left\{\overline{\alpha}_{ji}(f)\right\}$ and $\left\{\overline{n}_i(f)\right\}$. If $\overline{p}_1(f),...,\overline{p}_K(f)$ is not completely FDMA yet, do a flat FDMA reallocation to $\overline{p}_1(f),...,\overline{p}_K(f)$ in *every flat sub-channel* that has a flat frequency sharing of any subset of the $K$ inteferers. By Corollary 2.1, the resulting rate of user 0 satisfies $\overline{R}'_0 \geq \overline{R}_0 > R_0 - \varepsilon$. Finally, let $\varepsilon \to 0$. ∎

We conclude the section by combining Theorem 1 and Lemma 3 as follows:

*Theorem 2:* For any two users $i$ and $j$ (among all the $K$ users), for any frequency band $[f_1, f_2]$, if the normalized cross channel gains $\alpha_{ji}(f) \geq 1/2$ and $\alpha_{ij}(f) \geq 1/2$, $\forall f \in [f_1, f_2]$, then no matter from which user's point of view, an FDMA of user $i$ and user $j$ within this band is always preferred.

*Proof:* Suppose the spectrum and power allocation for user $i$ and $j$ are not FDMA, take a piecewise flat $\varepsilon$-approximation $\overline{\mathbf{p}}(f)$, $\left\{\overline{\alpha}_{ji}(f)\right\}$ and $\left\{\overline{n}_i(f)\right\}$, s.t. $\left|\overline{R}_k - R_k\right| < \varepsilon, \forall k = 1,...,K$. As in the proof of *Corollary 2.2*, with a flat FDMA reallocation of $\overline{p}_i(f)$ and $\overline{p}_j(f)$ in *every flat sub-channel* in $[f_1, f_2]$ that has a flat frequency sharing of user $i$ and $j$, Theorem 1 implies that user $i$ and $j$'s rates are increased or unchanged, and Lemma 3 implies that every one of the other $K-2$ users' rate is increased or unchanged. Finally, let $\varepsilon \to 0$. ∎

The pairwise condition $\alpha_{ji}(f) \geq 1/2$ and $\alpha_{ij}(f) \geq 1/2$ makes determining whether any two users should be orthogonally channelized depend only on the coupling conditions between the two of them. Furthermore, since this condition guarantees that an FDMA allocation between user $i$ and user $j$ benefits *every* one of the $K$ users, under this condition, *all the Pareto optimal points* of the rate region can be achieved with these two users having an FDMA allocation.

This pairwise condition is thus an example of distributed decision making (on whether to orthogonalize any pair of users) with optimality guarantees.

## IV. Optimal Spectrum Management in Two-User Symmetric Channels

In this section, we continue to analyze the optimal spectrum management in the cases with $\alpha(f) < 1/2$. We give a complete analysis of two-user (potentially frequency selective) symmetric Gaussian interference channels, defined as follows:

$$\begin{cases} \alpha_{12}(f) = \alpha_{21}(f) < 1/2, & \forall f \in [0,1], \\ n_1(f) = n_2(f), & \forall f \in [0,1]. \end{cases} \tag{4}$$

We choose the objective to be the *sum-rate* of the two users $R_1 + R_2$. General problems with $K \geq 2$ users and arbitrary weighted sum-rate objective functions in general asymmetric channels are discussed later in Section V.

Here, an *equal* power constraint

$$\int_0^1 p_i(f)df \leq p/2, \ i=1,2,$$

*or equivalently, a sum-power constraint*

$$\int_0^1 \big(p_1(f) + p_2(f)\big)df \leq p,$$

is assumed. (Equivalency is shown later in this section.) We begin with flat channels, and solve the optimal spectrum and power allocation by *computing a convex hull*. Based on this result, we show that finding the spectrum and power allocation that maximizes the non-concave sum-rate objective in symmetric frequency selective channels can be equivalently transformed into a convex optimization in the *primal* domain.

### A. Optimal Solutions for Flat Channels with a Sum-Power Constraint, or Equivalently, Equal Power Constraints

Consider a two-user symmetric flat Gaussian interference channel model:

$$\begin{cases} \alpha_{12}(f) = \alpha_{21}(f) = \alpha < 1/2, & \forall f \in [0,1], \\ n_1(f) = n_2(f) = n, & \forall f \in [0,1]. \end{cases}$$

WLOG, we can normalize the power and their constraints by the noise: $p_i(f) \leftarrow p_i(f)/n$, $p \leftarrow p/n$, and assume $n = 1$.

Firstly, we have the following theorem on the condition under which a flat FDMA scheme is better than a flat frequency sharing scheme. Denote $p_i$ to be the PSD of user $i = 1, 2$ in a flat frequency sharing scheme.

*Lemma 4:* For any flat frequency sharing power allocation, a flat FDMA power re-allocation (Figure 2) leads to a higher or unchanged sum-rate if and only if $p_1 + p_2 \geq 2\left(\dfrac{1}{2\alpha^2} - \dfrac{1}{\alpha}\right)$.

*Proof:* See Appendix C. ∎

Given the cross channel gains $\alpha$, Lemma 4 provides us a power region $P_{FDMA}$ within which flat FDMA has a higher sum-rate than flat frequency sharing, depicted as the shaded area in Figure 5 (with complement region $\overline{P}_{FDMA}$ also depicted). Clearly, if and only if $\alpha \geq 1/2$, $P_{FDMA}$ contains the entire non-negative quadrant. This provides a "weak" converse argument on the necessity of $1/2$ coupling condition derived in Section III, for FDMA to be *always optimal regardless of the power budget.*

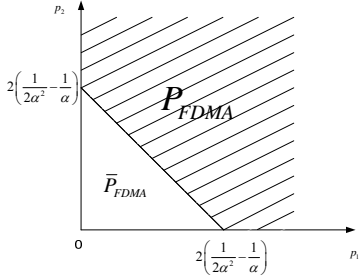

Fig. 5. The power region in which flat FDMA has higher sum-rate than flat frequency sharing.

Next, we derive the optimal flat frequency sharing scheme and the optimal flat FDMA scheme.

Denote the sum-rate of a flat frequency sharing by

$$f(p_1, p_2) = \log(1 + \frac{p_1}{1 + \alpha p_2}) + \log(1 + \frac{p_2}{1 + \alpha p_1}).$$

With a sum-power constraint $p_1 + p_2 \le p$ only, the maximum achievable sum-rate with flat frequency sharing, denoted by $f^*(p)$, is defined as the optimal value of the following optimization problem:

*Definition 5:*     $f^*(p) \triangleq \max \ f(p_1, p_2)$

$$s.t. \ p_1 + p_2 \le p$$
$$p_1 \ge 0, p_2 \ge 0 \quad (5)$$

Next, we show the form and the concavity of $f^*(p)$ in the region of $\overline{P}_{FDMA}$.

*Lemma 5:* When $0 < p \le 2\left(\frac{1}{2\alpha^2} - \frac{1}{\alpha}\right)$,

$$f^*(p) = 2\log\left(1 + \frac{p/2}{1 + \alpha p/2}\right) \quad (6)$$

is a concave function of the constraint $p$. The optimal flat frequency sharing scheme is $p_1 = p_2 = p/2$.

*Proof:* See Appendix C.  ∎

In comparison, we compute the maximum achievable sum-rate with a sum-power constraint for FDMA schemes, denoted by $h^*(p)$:

*Definition 6:* $h^*(p) = \max\limits_{p_1(f), p_2(f)} R_1 + R_2$

$$s.t. \int_0^1 (p_1(f) + p_2(f))df \le p,$$
$$p_1(f)p_2(f) = 0, \ p_1(f) \ge 0, p_2(f) \ge 0, \ \forall f \in [0,1],$$
$$R_1 = \int_0^1 \log(1 + p_1(f))df, \ R_2 = \int_0^1 \log(1 + p_2(f))df.$$

From FDMA and the symmetry assumption of the channel, the sum-rate of both users is equivalent to the rate of a single user with a power constraint of $p$. With the water-filling principle, $h^*(p)$ is achieved when the PSD over the whole band is flat. In other words, both users' powers are allocated mutually non-overlapped and collectively filling the whole band uniformly. Accordingly, we have the following lemma.

*Lemma 6:* The maximum achievable sum-rate with FDMA

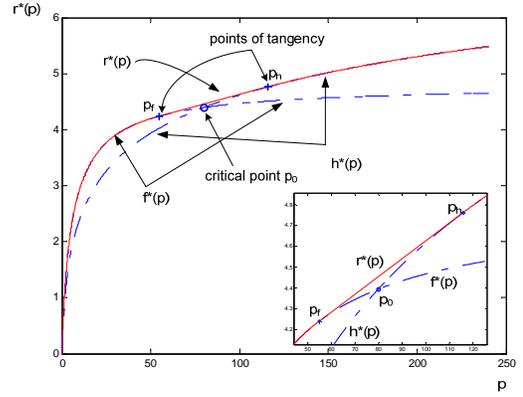

Fig. 6. The maximum achievable rate as the convex hull of the rates of flat FDMA and flat frequency sharing.

is

$$h^*(p) = \log(1 + p). \quad (7)$$

*Define the critical point* $p_0 = 2\left(\frac{1}{2\alpha^2} - \frac{1}{\alpha}\right)$. Directly from Lemma 4, it can be verified that $f^*(p_0) = h^*(p_0)$.

As $f^*(p)$ and $h^*(p)$ are both increasing and concave, the upper envelope of $f^*(p)$ and $h^*(p)$ is given by

$$r(p) \triangleq \max\{f^*(p), h^*(p)\} = \begin{cases} f^*(p), & p \in [0, p_0] \\ h^*(p), & p \in [p_0, \infty) \end{cases}.$$

Furthermore, as $0 < \alpha < 1/2$,

$$\frac{d}{dp}f^*(p)\bigg|_{p=p_0} = \frac{4\alpha^3}{1-\alpha} < \frac{\alpha^2}{(1-\alpha)^2} = \frac{d}{dp}h^*(p)\bigg|_{p=p_0}, \quad (8)$$

and the upper envelope $r(p)$ is non-concave in $[0, \infty)$.

Next, $r^*(p)$ *is defined to be the unique convex hull of* $r(p)$:

$$r^*(p) \triangleq conv_p(r(p)).$$

A typical plot of $f^*(p)$, $h^*(p)$, and their upper envelope convex hull $r^*(p)$ is given in Figure 6. Since $f^*(p)$ and $h^*(p)$ are themselves concave, the convex hull of the upper envelope is found by computing their *common tangent line*. For example, In Figure 6, $\alpha$ is chosen to be 0.1. $f^*(p)$ and $h^*(p)$ intersects at $p_0 = 2\left(\frac{1}{2\alpha^2} - \frac{1}{\alpha}\right) = 80$. The two points of tangency are $p_f = 54.931$, $p_h = 115.938$.

In order to find the common tangent line of $f^*(p)$ and $h^*(p)$, the two points of tangency $p_f$ and $p_h$ are determined by

$$\frac{d}{dp}f^*(p)\bigg|_{p=p_f} = \frac{d}{dp}h^*(p)\bigg|_{p=p_h} = \frac{h^*(p_h) - f^*(p_f)}{p_h - p_f},$$

which simplifies to finding $p_f$ by solving

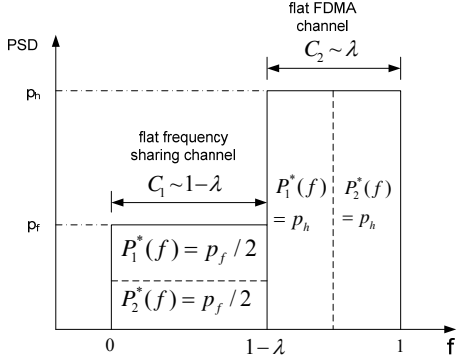

Fig. 7. The optimal spectrum and power allocation as a mixture of flat FDMA and flat frequency sharing.

$$\frac{p_f\left(\alpha(1+\alpha)p_f + 4\alpha - 2\right)}{\left(\alpha p_f + 2\right)\left((1+\alpha)p_f + 2\right)} = \log\left(\frac{(\alpha p_f + 2)^3}{4\left((1+\alpha)p_f + 2\right)}\right), \quad (9)$$

and computing $p_h$ by

$$p_h = \frac{1}{4}p_f\left(\alpha(1+\alpha)p_f + 4\alpha + 2\right).$$

$p_f$ and $p_h$ can be obtained by solving the closed form equation (9) where various numerical methods can be applied. From many numerical examples, we observed that (9) always has one valid fix point solution.

Next, we provide the main theorem of this sub-section.

*Definition 7:* In a flat symmetric Gaussian interference channel with $\alpha < 1/2$, define $r^o(p)$ to be maximum achievable sum-rate with a sum-power constraint $p$:

$$r^o(p) \triangleq \max_{\mathbf{p}(f)} \int_0^1 r_1(\mathbf{p}(f), f) + r_2(\mathbf{p}(f), f) df \qquad (10)$$

$$s.t. \ \int_0^1 \left(p_1(f) + p_2(f)\right) df \le p,$$

$$p_1(f) \ge 0, p_2(f) \ge 0, \ \forall f \in (f_1, f_2),$$

$$r_1(\mathbf{p}(f), f) = \log\left(1 + \frac{p_1(f)}{1 + \alpha p_2(f)}\right),$$

$$r_2(\mathbf{p}(f), f) = \log\left(1 + \frac{P_2(f)}{1 + \alpha P_1(f)}\right).$$

*Theorem 3:*

$$r^o(p) = r^*(p).$$

While the proof of the achievability of $r^*(p)$ is fairly straightforward, the proof of the converse follows from Jensen's inequality, as we recognize that *all* allocation schemes $\mathbf{p}(f)$ are *pointwise* either flat frequency sharing or flat FDMA.

*Proof of Theorem 3:*

i) $r^*(p) \le r^o(p)$ (Achievability of $r^*(p)$).

The achievability of $r^*(p)$ when $0 < p \le p_f$ or $p \ge p_h$ is immediate. When $p_f \le p \le p_h$,

$$r^*(p) = f^*(p_f) + \lambda\left(h^*(p_h) - f^*(p_f)\right),$$

where $\lambda = \dfrac{p - p_f}{p_h - p_f}$, and $r^*(p)$ is achievable by the following scheme as depicted in Figure 7: The band of the original channel is split into two orthogonal channels: $C_1$ with bandwidth $1 - \lambda$, and $C_2$ with bandwidth $\lambda$.

In $C_1$, a flat frequency sharing with a PSD of $p_f/2$ for each user is applied, achieving a sum-rate of $f^*_{C_1} = (1 - \lambda)f^*(p_f)$.

In $C_2$, a flat FDMA with a PSD of $p_h$ for each user is applied, achieving a sum-rate of $h^*_{C_2} = \lambda h^*(p_h)$.

Note that the sum-power constraint is satisfied by such a combination of flat frequency sharing and flat FDMA:

$$(1 - \lambda)p_f + \lambda p_h = p.$$

Therefore, the sum-rate

$$f^*_{C_1} + h^*_{C_2} = (1 - \lambda)f^*(p_f) + \lambda h^*(p_h) = r^*(p)$$

can be achieved in the original problem (10).

ii) $r^o(p) \le r^*(p)$ (Converse)

For any given $p$, let $\left\{p_1^o(f), p_2^o(f)\right\}$ be an optimal scheme that achieves $r^o(p)$. Define the sum-rate density

$$r_p^o(f) \triangleq \log\left(1 + \frac{p_1^o(f)}{1 + \alpha p_2^o(f)}\right) + \log\left(1 + \frac{p_2^o(f)}{1 + \alpha p_1^o(f)}\right),$$

and $p^o(f) \triangleq p_1^o(f) + p_2^o(f)$. Clearly, $r^o(p) = \int_0^1 r_p^o(f) df$.

From Lemma 5, when $p^o(f) \le 2\left(\dfrac{1}{2\alpha(f)^2} - \dfrac{1}{\alpha(f)}\right)$,

$$r_p^o(f) \le f^*(p^o(f)).$$

From Lemma 4 and 6, when $p^o(f) > 2\left(\dfrac{1}{2\alpha(f)^2} - \dfrac{1}{\alpha(f)}\right)$,

$$r_p^o(f) \le h^*(p^o(f)).$$

Thus, $r_p^o(f) \le \max\left\{f^*(p^o(f)), h^*(p^o(f))\right\} \le r^*(p^o(f))$, and

$$r^o(p) = \int_0^1 r_p^o(f) df \le \int_0^1 r^*(p^o(f)) df$$

$$\le r^*\left(\int_0^1 p^o(f) df\right) \le r^*(p).$$

The second inequality arises from the concavity of $r^*(p)$ and Jensen's inequality, and the last inequality arises from the sum-power constraint and the fact that $r^*(p)$ is increasing. ∎

The mixture of a flat frequency sharing and a flat FDMA shown in Figure 7 represents the general form of the optimal spectrum and power allocation achieving $r^*(p)$.

The computation of the optimal spectrum management scheme is summarized in Procedure 1. Note that there always exists an optimal spectrum and power allocation with two users each using the same total power of $p/2$. Therefore, the above optimal solution with a *sum*-power constraint directly leads to

---

**Procedure 1**

**Computing the optimal spectrum management for two-user symmetric flat channels**

1) Solve the two points of tangency $p_f$ and $p_h$ of the convex hull of $f^*(x)$ and $h^*(x)$ :

a. Solve equation (9) numerically to find $p_f$

$$\frac{p_f\left(\alpha(1+\alpha)p_f+4\alpha-2\right)}{\left(\alpha p_f+2\right)\left((1+\alpha)p_f+2\right)}=\log\left(\frac{(\alpha p_f+2)^3}{4\left((1+\alpha)p_f+2\right)}\right).$$

b. Compute $p_h$ by $p_h=\frac{1}{4}\left(\alpha(1+\alpha)p_f+4\alpha+2\right)$.

2) Compute the maximum achievable sum-rate $r^*(p)$ :

If $p \le p_f$, $r^*(p) = f^*(p)$ .

　　　　Allocate $p_1(f)=p_2(f)=p/2,\ \forall f$ .

If $p \ge p_h$, $r^*(p)=h^*(p)$ .

　　　　Allocate $p_1(f)$ and $p_2(f)$ such that

$$\begin{cases} p_1(f)p_2(f)=0,\ \forall f \\ p_1(f)+p_2(f)=p,\ \forall f \end{cases}.$$

If $p_f < p < p_h$ ,

$$r^*(p)=f^*(p_f)+\frac{h^*(p_h)-f^*(p_f)}{p_h-p_f}\left(p-p_f\right).$$

　　a. Compute $\lambda=\dfrac{p-p_f}{p_h-p_f}$

　　b. Separate [0,1] into two disjoint channels: $C_1$ with bandwidth $1-\lambda$, $C_2$ with bandwidth $\lambda$ .

　　c. Allocate power as follows (Figure 7):
　　　　In $C_1$, $p_1(f)=p_2(f)=p_f/2$ .

　　　　In $C_2$, $\begin{cases} p_1(f)p_2(f)=0,\ \forall f \\ p_1(f)+p_2(f)=p_h,\ \forall f \end{cases}$

---

the optimal solution with *equal* individual power constraints:

*Corollary 3:* In a flat symmetric Gaussian interference channel with $\alpha < 1/2$ , the maximum sum-rate defined as the optimal value of the following optimization problem

$$\max_{\mathbf{p}(f)} \int_0^1 r_1(\mathbf{p}(f),f)+r_2(\mathbf{p}(f),f)df$$

$$s.t.\ \int_0^1 p_i(f)df \le p/2,\ i=1,2,$$
$$p_1(f)\ge 0, p_2(f)\ge 0,\ \forall f\in[0,1].$$

is $r^*(p)$ .

*Proof:* On the one hand, the equal power constraints imply the sum-power constraint. On the other hand, the optimal value with the sum-power constraint can be achieved with the equal power constraints. ∎

## B. Generalizations to Frequency Selective Channels

In this sub-section, we extend the sum-rate maximization problem to the symmetric frequency selective Gaussian interference channel.

$$\begin{cases} \alpha_{12}(f)=\alpha_{21}(f)=\alpha(f),\ \ \forall f\in[0,1], \\ n_1(f)=n_2(f)=n(f),\ \ \forall f\in[0,1]. \end{cases}$$

With

$$r_1(\mathbf{p}(f),f)=\log\left(1+\frac{p_1(f)}{n(f)+p_2(f)\alpha(f)}\right),$$

$$r_2(\mathbf{p}(f),f)=\log\left(1+\frac{p_2(f)}{n(f)+p_1(f)\alpha(f)}\right),$$

define $r^o$ to be the maximum achievable sum-rate with a sum-power constraint as follows:

*Definition 8:* $r^o \triangleq \max_{\mathbf{p}(f)} \int_0^1 r_1(\mathbf{p}(f),f)+r_2(\mathbf{p}(f),f)df$ 　　(11)

$$s.t.\ \int_0^1\left(p_1(f)+p_2(f)\right)df \le p,$$
$$p_1(f)\ge 0, p_2(f)\ge 0,\ \forall f\in[0,1].$$

Note that the objective function *is separable in f.* (The whole problem is, of course, not immediately separable in $f$ because of the total power constraint across the whole band.)

*Remark 3:* Because *for every fixed* $f\in[0,1]$, $r_1(\mathbf{p}(f),f)+r_2(\mathbf{p}(f),f)$ is *non-concave* in $\{p_1(f),p_2(f)\}$ , the above infinite dimensional problem is a *non-convex optimization*.

Next, we derive a *primal domain convex relaxation* of (11). We first normalize the PSD and the sum-PSD by $n(f)$ :

*Definition 9: At every frequency* $f\in[0,1]$,

1. $\tilde{p}_1(f)\triangleq\dfrac{\tilde{p}_1(f)}{n(f)}, \tilde{p}_2(f)\triangleq\dfrac{\tilde{p}_2(f)}{n(f)}, \tilde{p}(f)\triangleq\tilde{p}_1(f)+\tilde{p}_2(f)$ .

2. In the same form of (6) and (7) with $\alpha(f)$ instead of $\alpha$ :

$$f^*(p,f)\triangleq 2\log\left(1+\frac{p/2}{1+\alpha(f)p/2}\right), h^*(p,f)\triangleq\log(1+p),\text{ and}$$

$$r^*(p,f)\triangleq conv_p\left(\max\left\{f^*(p,f),h^*(p,f)\right\}\right).$$

Note that the convex hull operation is done *along the power dimension for every fixed f, (not along the frequency dimension.)* $\forall f\in[0,1]$, $p_f(f)$, $p_h(f)$, and $r^*(p,f)$ are computed in the same way as in Procedure 1 with $\alpha(f)$ instead of $\alpha$ .

In the (separable) objective function of (11), *at every frequency f,* we *replace the non-concave* $r_1(\mathbf{p}(f),f)+r_2(\mathbf{p}(f),f)$ *with the concave* $r^*(\tilde{p}(f),f)$ (concave in the first variable $\tilde{p}(f)$ ), and define $r^*$ to be the corresponding maximum achievable value as follows:

*Definition 10:* $r^* \triangleq \max_{\tilde{p}(f)} \int_0^1 r^*(\tilde{p}(f),f)df$ 　　(12)

$$s.t.\ \int_0^1\tilde{p}(f)n(f)df \le p,\ \tilde{p}(f)\ge 0, \forall f\in[0,1].$$

*Remark 4:* For every fixed $f\in[0,1]$ , $r^*(\tilde{p}(f),f)$ is *concave in* $\tilde{p}(f)$ . The constraint is *linear in* $\tilde{p}(f), \forall f$ . Thus, the above infinite dimensional problem is a *convex optimization*.

Now, we have the following theorem:

*Theorem 4:*

$$r^o = r^*.$$

The proof of the converse is similar to that in Theorem 3. For the proof of the achievability of $r^*$, as the channel is frequency selective, we need to introduce a piecewise flat $\varepsilon$ - approximation, and the remaining proof exactly follows that in Theorem 3.

*Proof of Theorem 4:*

*i)* $r^o \leq r^*$ (Converse).

It is sufficient to prove the inequality between the integrands in (11) and (12). As in the proof of Theorem 3, from Lemma 4, 5, 6,

$$r_1(\mathbf{p}(f), f) + r_2(\mathbf{p}(f), f)$$

$$= \log\left(1 + \frac{\tilde{p}_1(f)}{1 + \tilde{p}_2(f)\alpha(f)}\right) + \log\left(1 + \frac{\tilde{p}_2(f)}{1 + \tilde{p}_1(f)\alpha(f)}\right)$$

$$\leq \max\left\{f^*(\tilde{p}(f), f), h^*(\tilde{p}(f), f)\right\} \leq r^*(\tilde{p}(f), f).$$

*ii)* $r^* \leq r^o$ (Achievability).

Let sum-PSD $\vec{p}^*(f)$ be an optimal solution of (12) such that $\int_0^1 r^*(\vec{p}^*(f), f)df = r^*$. Then, $\forall \varepsilon > 0$:

By Lemma 1, based on $\vec{p}^*(f)$ and $\{\alpha_{ji}(f)\}$, take a piecewise flat $\varepsilon$ - approximation $\overline{p}^*(f)$ and $\{\overline{\alpha}_{ji}(f)\}$, s.t.

$$\left|\int_0^1 \overline{r}^*(\overline{p}^*(f), f)df - r^*\right| < \varepsilon,$$

where $\overline{p}^*(f)$ is a piecewise flat *sum-PSD*, and $\overline{r}^*(\overline{p}^*(f), f)$ is computed with $\overline{p}^*(f)$ and $\{\overline{\alpha}_{ji}(f)\}$. (Note that, since the noise PSD is already normalized to 1 as in Definition 9, no further piecewise flat approximation of the noise is needed.)

Based on the piecewise flat $\varepsilon$ - approximation, *in every flat sub-channel with a flat* $\overline{p}^*(f)$, as in the proof of Theorem 3, $\overline{r}^*(\overline{p}^*(f), f)$ can be achieved by further dividing this flat sub-channel into two sub-bands, applying a flat frequency sharing and a flat FDMA respectively (c.f. Figure 7). Removing the normalization by multiplying by $n(f)$, denote the resulting allocation scheme by $\mathbf{p}^o(f) = \left[p_1^o(f), p_2^o(f)\right]^T$, achieving the same sum-rate

$$\int_0^1 \overline{r}_1(\mathbf{p}^o(f), f) + \overline{r}_2(\mathbf{p}^o(f), f)df = \int_0^1 \overline{r}^*(\overline{p}^*(f), f)df,$$

where $\overline{r}_1(\mathbf{p}^o(f), f)$ and $\overline{r}_2(\mathbf{p}^o(f), f)$ are computed with the piecewise flat approximate channel responses $\{\overline{\alpha}_{ji}(f)\}$.

Then,

$$r^o \geq \int_0^1 r_1(\mathbf{p}^o(f), f) + r_2(\mathbf{p}^o(f), f)df$$

$$\geq \int_0^1 \overline{r}_1(\mathbf{p}^o(f), f) + \overline{r}_2(\mathbf{p}^o(f), f)df$$

$$= \int_0^1 \overline{r}^*(\overline{p}^*(f), f)df > r^* - \varepsilon,$$

where the first inequality occurs because $\mathbf{p}^o(f)$ is a *feasible* solution of (11); the second inequality arises because (by P2 from Lemma 1) $\{\overline{\alpha}_{ji}(f)\} \geq \{\alpha_{ji}(f)\}$, $\forall i, j, \forall f$, i.e. the $\varepsilon$ - approximation worsens the channel responses, resulting in lower rates.

Finally, let $\varepsilon \to 0$. ∎

Therefore, although the integrand in (12) is a direct convex relaxation of that in (11), the *optimal objective value* of the problem does not change, and the original non-convex optimization (11) is equivalently transformed to the convex optimization (12).

Finally, for the same reasons as in part A, the optimal solution with equal individual power constraints is the same as that with a corresponding sum-power constraint.

*Remark 5:* Throughout this sub-section, we have worked with a sum-power constraint to gain brevity in derivations of the results for the fully symmetric cases. One may also derive the results directly with equal individual power constraints. In Section V, as we consider potentially *asymmetric* channels, we will directly work with individual power constraints.

## V. Optimal Spectrum Management in the General Cases

In Section IV, we solved the sum-rate maximization problem in two-user symmetric frequency selective channels with equal power (or sum-power) constraints. In this section, we make the following generalizations:

1. Two-user → *K*-user,
2. Equal power constraints → arbitrary individual power constraints,
3. Symmetric channel → arbitrary asymmetric channel,
4. Sum-rate → arbitrary weighted sum-rate.

The general optimization problem is thus the following:

$$\max_{\mathbf{p}(f)} \int_0^1 \sum_{i=1}^{K} w_i r_i(\mathbf{p}(f), f)df \qquad (13)$$

$$s.t. \int_0^1 \mathbf{p}(f)df \leq \mathbf{p}, \quad \mathbf{p}(f) \geq 0, \ \forall f \in [0,1],$$

$$r_i(\mathbf{p}(f), f) = \log\left(1 + \frac{p_i(f)}{n_i(f) + \sum_{j \neq i} p_j(f)\alpha_{ji}(f)}\right).$$

Next, we analyze this general problem in parallel with the analysis in Section IV, and show that the same basic ideas in Section IV generalize here.

### A. Optimal Solutions for Flat Channels

Consider a *K*-user (potentially asymmetric) flat channel:

$$\alpha_{ji}(f) = \alpha_{ji}, \ n_i(f) = n_i, \forall f \in [0,1], \forall i, j.$$

First, consider the weighted sum-rate achieved with *flat*

power allocations $\mathbf{p}(f) = \mathbf{P}, \forall f \in [0,1]$, defined as

$$R(\mathbf{P}) \triangleq \sum_{i=1}^{K} w_i \log \left( 1 + \frac{P_i}{n_i + \sum_{j \neq i} P_j \alpha_{ji}} \right). \quad (14)$$

Denote its $K$ dimensional convex hull function by

$$R^*(\mathbf{P}) \triangleq conv_{\mathbf{P}} (R(\mathbf{P})). \quad (15)$$

The original problem (13) in flat channels can be rewriten as

*Definition 11:* $R^o(\mathbf{p}) \triangleq \max_{\mathbf{p}(f)} \int_0^1 R(\mathbf{p}(f)) df$

$$s.t. \int_0^1 \mathbf{p}(f) df \leq \mathbf{p}, \ \mathbf{p}(f) \geq 0, \ \forall f \in [0,1].$$

Now, we have the following theorem,

*Theorem 5:*

$$R^o(\mathbf{p}) = R^*(\mathbf{p}),$$

and the optimal spectrum and power allocation $\mathbf{p}^o(f)$ consists of $K+1$ sub-bands, with $\mathbf{p}^o(f)$ flat in each of the sub-bands.

*Proof:*

The proof is in parallel with that of Theorem 3.

1. $R^*(\mathbf{p}) \leq R^o(\mathbf{p})$ (Achievability).

As $R^*(\mathbf{p}) = conv_{\mathbf{P}} (R(\mathbf{p}))$, by Carathéodory's theorem,

$$\exists \sum_{k=1}^{K+1} c^{(k)} = 1, \ \sum_{k=1}^{K+1} c^{(k)} \mathbf{p}^{(k)} = \mathbf{p}, \ c^{(k)} \geq 0, \ s.t.$$

$$R^*(\mathbf{p}) = \sum_{k=1}^{K+1} c^{(k)} R(\mathbf{p}^{(k)}).$$

Accordingly, we can divide the band $[0,1]$ into $K+1$ sub-bands, each with a bandwidth of $c^{(k)}$ and uses the flat power levels of $\mathbf{p}^{(k)} = [p_1^{(k)}, ..., p_K^{(k)}]^T$ for the $K$ users.

2. $R^o(\mathbf{p}) \leq R^*(\mathbf{p})$ (Converse).

For any feasible allocation scheme $\mathbf{p}(f), f \in [0,1]$.

$$\int_0^1 R(\mathbf{p}(f)) df \leq \int_0^1 R^*(\mathbf{p}(f)) df \leq R^* \left( \int_0^1 \mathbf{p}(f) df \right) \leq R^*(\mathbf{p}),$$

where the first inequality is from definition (15), the second inequality arises from Jensen's inequality, and the third inequality arises from the fact that $R^*(\mathbf{P})$ is increasing in $\mathbf{P}$. ∎

*Remark 6:* In the literature, it was first shown that allocation schemes consisting of $2K$ sub-bands of *flat* allocations are sufficient to achieve any Pareto optimality [9], and this sufficient number of sub-bands was later refined to $K+1$ [17]. From Theorem 5, the sufficiency of $K+1$ sub-bands is also immediately implied by the fact that the optimal value and solution are obtained by nothing more than computing the convex hull (15) of a non-concave function (14).

## B. Generalizations to Frequency Selective Channels

In frequency selective channels, define the *weighted sum-rate density function* as

$$R(\mathbf{P}, f) \triangleq \sum_{i=1}^{K} w_i r_i (\mathbf{P}, f) \quad (16)$$

Problem (13) can then be rewritten as the following:

*Definition 12:* $R^o \triangleq \max_{\mathbf{p}(f)} \int_0^1 R(\mathbf{p}(f), f) df \quad (17)$

$$s.t. \int_0^1 \mathbf{p}(f) df \leq \mathbf{p}, \ \mathbf{p}(f) \geq 0, \ \forall f \in [0,1].$$

Clearly, for every fixed $f \in [0,1]$, $R(\mathbf{p}(f), f)$ is non-concave in $\mathbf{p}(f)$, and the (17) is an infinite dimensional non-convex optimization.

At every frequency $f \in [0,1]$, define

$$R^*(\mathbf{P}, f) \triangleq conv_{\mathbf{P}} R(\mathbf{P}, f),$$

i.e., *the convex hull of $R(\mathbf{P}, f)$ along the $K$ dimensions of power $\mathbf{P}$*. Note that the convex hull operation is *not* taken along the frequency dimension *f*. (Note that $R^*(\mathbf{P}, f)$ is concave in $\mathbf{P}$ for every fixed *f*, but not necessarily jointly concave in $\mathbf{P}, f$.)

Next, we derive the following primal domain convex relaxation of (17): *At every frequency f, we replace the non-concave $R(\mathbf{p}(f), f)$ with the concave $R^*(\mathbf{p}(f), f)$* (concave in the first variable $\mathbf{p}(f)$), and define $R^*$ to be the corresponding maximum achievable value as follows:

*Definition 13:* $R^* \triangleq \max_{\mathbf{p}(f)} \int_0^1 R^*(\mathbf{p}(f), f) df \quad (18)$

$$s.t. \int_0^1 \mathbf{p}(f) df \leq \mathbf{p}, \ \mathbf{p}(f) \geq 0, \ \forall f \in [0,1].$$

Clearly, (18) is an infinite dimensional convex optimization, because $\forall f \in [0,1]$, the integrand is a concave function of the variables $\{\mathbf{p}(f)\}$, and the constraint is linear in $\{\mathbf{p}(f)\}$.

Now, we have the following theorem

*Theorem 6:*

$$R^o = R^*.$$

*Proof:* The detailed proof is relegated to Appendix C, in parallel with the proof of Theorem 4. ∎

Therefore, the optimal value for the non-convex optimization (17) *equals* that of its convex relaxation (18).

## C. On the Complexity of Solving the General Problem

For general piecewise bounded continuous channel responses, problem (17) can have up to an uncountably infinite number of dimensions, for which describing the complexity of solving the continuous frequency optimal solution is pointless. However, one can first *approximate* the channel responses by *piecewise flat* functions of frequency, which is also the way by which spectrum management problems are approached in practice.

With piecewise flat channel responses, denote the corresponding flat sub-channels by $I_1, I_2, ..., I_M$, each with bandwidth $b_m$. (Note that the channels $\{I_m\}$ are viewed as *given*, and their bandwidths $\{b_m\}$ are *not* variables to optimize.) One can consider two types of problems distinguished by the assumptions on power allocations:

*Case a.* $\mathbf{p}(f)$ is piecewise bounded continuous functions.

*Case b.* $\mathbf{p}(f)$ must be *flat in every flat sub-channel $I_m$*.

For example, consider a single flat band. It makes a fundamental difference whether we allow a user to subdivide this flat band and use different PSD in different sub-bands. If so, it is Case *a.*, and the problem model is still *continuous* frequency; Otherwise, it is Case *b.*, and it corresponds to the *discrete* frequency model.

It has been proven that finding the optimal solution with the *discrete* frequency model (Case *b.*) is NP hard [14]. This is *not inconsistent* with the convex formulations for the *continuous* frequency model in this paper, because the assumptions made on power allocation are different (Case *a.* vs. Case *b.*)

Next, we discuss the complexity of solving the *continuous* frequency optimal spectrum management (Case *a.*) in *piecewise flat channels*. From Theorem 6, it is sufficient to solve the convex optimization (18), which consists of two general steps:

*Step 1)* Compute the convex hull function $R^*(\mathbf{P}, f)$ *at every frequency* $f \in [0,1]$.

*Step 2)* Optimize $\mathbf{p}(f)$ with the objective $\int_0^1 R^*(\mathbf{p}(f), f) df$.

In Step 1, given the channel parameters for each flat sub-channel $I_m$, $m = 1,...,M$, a convex hull function $R_m^*(\mathbf{P})$ $\left( \triangleq R^*(\mathbf{P}, f), f \in I_m, \right)$ $\forall \mathbf{P} \geq 0$ is computed. Numerically and approximately computing a convex hull is itself a broad and important topic (see e.g. [25]). This computational issue is not further discussed in this paper, and is left for future work. We note that it remains unclear whether this computational issue is easier to deal with than the NP hardness in the discrete frequency model.

In Step 2, given the convex hull functions for all the flat sub-channels, as the number of sub-channels $M$ is *finite*, problem (18) becomes *finite dimensional*, with an *increasing concave* utility function $R_m^*(\mathbf{p}(f))$ in each sub-channel $I_m$. Now, because each $I_m$ has *flat* channel parameters and $R_m^*(\mathbf{P})$ is increasing concave, by Jensen's inequality, the optimal solution must satisfy that $\mathbf{p}(f)$ is *flat* in *each* sub-channel $I_m$, i.e., $\forall m = 1,...,M$, $\exists \mathbf{p}(m) \geq 0$, *s.t.* $\mathbf{p}(f) = \mathbf{p}(m)$, $\forall f \in I_m$. Problem (18) then becomes

$$\max_{\mathbf{p}(m)} \sum_{m=1}^{M} b_m \cdot R^*(\mathbf{p}(m))$$
$$s.t. \sum_{m=1}^{M} b_m \cdot \mathbf{p}(m) \leq \mathbf{p}, \ \mathbf{p}(m) \geq 0, \ \forall m = 1,2,...,M. \quad (19)$$

(Recall that $b_m$ $(m = 1,...,M)$ is the bandwidth of sub-channel $I_m$, and is *not* an optimization variable). Problem (19) is a classic *convex* optimization that has efficient polynomial time algorithms to solve the global optimal solution. (For example, a standard dual decomposition algorithm works, see e.g. [7] among many others.)

In summary, the critical complexity in solving the general problem (17) based on Theorem 6 lies in computing approximate convex hull functions. While computing convex hull functions given channel parameters may be

computationally costly, this two-step method does have the following advantage:

*Corollary 4*: Once the channel parameters are given, the $M$ convex hull functions $R_m^*(\mathbf{P})$ are computed *for one time*. Then, no matter how the power constraints may vary due to problem needs, the additional complexity cost of solving the optimal solution (Step 2) is only polynomial time.

This separation of the complexity in dealing with channel responses and power constraints does *not* appear in the *discrete* frequency model, due to the fundamental difference between the power allocation assumption of the continuous frequency model and that of the discrete frequency model (Case *a.* vs. Case *b.*) For the discrete frequency model, the constraint that a user must use a *flat PSD within every (flat) sub-channel* leads to the well known NP hardness. In contrast, for the continuous frequency model, the main complexity is from computing convex hull functions.

### D. On the Zero Duality Gap

It has been proved that the continuous frequency non-convex optimization (17) has an *exact* zero duality gap [14] [21]. It is pointed out that the zero duality gap comes from a *time sharing condition* [21]. It is also proved using the nonatomic property of the Lebesgue measure [14].

We show that this is also immediately implied by Theorem 6.

*Definition 14:*

For problem (17), its Lagrange dual is defined as

$$L(\mathbf{p}(f), \boldsymbol{\lambda}) \triangleq \int_0^1 R(\mathbf{p}(f), f) df - \boldsymbol{\lambda}^T \left( \int_0^1 \mathbf{p}(f) df - \mathbf{p} \right).$$

Its dual objective and dual optimal value are defined as

$$g(\boldsymbol{\lambda}) \triangleq \sup_{\mathbf{p}(f) \geq 0} L(\mathbf{p}(f), \boldsymbol{\lambda}), \text{ and } D^o \triangleq \min_{\boldsymbol{\lambda} \geq 0} g(\boldsymbol{\lambda}).$$

Similarly, for problem (18), its Lagrange dual, dual objective, and dual optimal value are defined as

$$\hat{L}(\mathbf{p}(f), \boldsymbol{\lambda}) \triangleq \int_{f_1}^{f_2} R^*(\mathbf{p}(f), f) df - \boldsymbol{\lambda}^T \left( \int_{f_1}^{f_2} \mathbf{p}(f) df - \mathbf{p} \right).$$

$$\hat{g}(\boldsymbol{\lambda}) \triangleq \sup_{\mathbf{p}(f) \geq 0} \hat{L}(\mathbf{p}(f), \boldsymbol{\lambda}), \text{ and } D^* \triangleq \min_{\boldsymbol{\lambda} \geq 0} \hat{g}(\boldsymbol{\lambda}).$$

*Corollary 5:* The non-convex optimization (17) has a zero duality gap.

*Proof:*

Since $R^*(\mathbf{P}, f) \geq R(\mathbf{P}, f)$, $\forall \mathbf{P}, \forall f$, we have

$$\hat{L}(\mathbf{p}(f), \boldsymbol{\lambda}) \geq L(\mathbf{p}(f), \boldsymbol{\lambda}) \Rightarrow \hat{g}(\boldsymbol{\lambda}) \geq g(\boldsymbol{\lambda}), \ \forall \boldsymbol{\lambda} \geq 0$$
$$\Rightarrow D^* \geq D^o.$$

Note that the primal optimal values for (17) and (18) are $R^o$ and $R^*$. Therefore,

$$R^* = D^* \geq D^o \geq R^o = R^* \Rightarrow D^o = R^o,$$

where the first equality occurs because problem (18) is a convex optimization and has *strong* duality [3]; the second inequality is from the *weak* duality of the non-convex optimization (17); the key step is the second equality $R^o = R^*$ from Theorem 6. ∎

Furthermore, it has been shown that, under mild technical conditions, the non-convex optimization for the *discrete* frequency model has an *asymptotically* zero duality gap as the number of sub-channels goes to infinity [21]. The result is rigorously generalized to include Lebesgue integrable PSDs in [14]. Indeed, for a piecewise bounded continuous frequency channel, as it is divided into more and finer/flatter sub-channels, the difference between the power allocation assumptions Case *a.* and Case *b.* vanishes (discrete frequency model → continuous frequency model.) The intuition is that we can *bundle a large number of similar flat* sub-channels, treat them as *one combined flat channel*, compute the *continuous* frequency power allocation, and accordingly distribute the power within these roughly identical sub-channels (as a discrete approximation of the continuous allocation.)

## VI. CONCLUSIONS

In this paper, we considered two general problems for continuous frequency optimal spectrum management in Gaussian interference channels: *1)* The channel conditions under which FDMA schemes are Pareto optimal; *2)* Equivalent convex formulations for the non-convex weighted sum-rate maximization problem.

Firstly, we have shown that for any two (among *K*) users, as long as the two normalized cross channel gains between them are both larger than or equal to 1/2, an FDMA allocation between these two users *benefits every one of the K users*. Therefore, under this pairwise condition, any Pareto optimal point of the *K*-user rate region can be achieved with this pair of users using orthogonal channels. The pairwise nature of the condition allows a completely distributed decision on whether any two users should use orthogonal channels, without loss of any Pareto optimality.

Next, we have shown that the classic non-convex *weighted* sum-rate maximization in *K-user asymmetric frequency selective* channels *can be equivalently transformed in the primal domain to a convex optimization*. We first analyzed in detail the sum-rate maximization in two-user symmetric flat channels, and showed that the optimal solution consists of one sub-band of flat frequency sharing, and one sub-band of flat FDMA. We generalized the results to weighted sum-rate maximization in *K*-user asymmetric flat channels: we showed that the optimal value is computed as the *convex hull* of the non-concave objective function, and the piecewise flat optimal solution is obtained based on the convex combination used in computing the point on the convex hull. Finally, a primal domain convex formulation is established for frequency selective channels. For piecewise flat channels, we showed that the main complexity lies in computing convex hull functions.

This paper tries to provide a unified and in-depth view on solving the optimal spectrum management problem for the *continuous* frequency model. The multi-channel *discrete* frequency model is fundamentally different from (although related to) the continuous frequency model (even with piecewise flat channel responses). As problems with the discrete frequency model are in general NP-compete, finding practical algorithms to find approximately optimal solutions has attracted many research endeavors, and continues to be very interesting.

## APPENDIX A

*Proof of Lemma 1:*

First, we prove for the case that $\mathbf{p}(f), \{\alpha_{ji}(f)\}$ and $\{n_i(f)\}$ are *bounded continuous* in $[0,1]$, (*not piecewise*.) It is then immediate to generalize to *piecewise* bounded continuous functions with a *finite* number of discontinuities.

*1)* Since $U(\mathbf{p}, \boldsymbol{\alpha}, \mathbf{n})$ is a uniformly continuous function of $\{\{p_i\}, \{\alpha_{ji}\}, \{n_i\}\}$, $\forall \varepsilon > 0, \exists \varepsilon' > 0$, s.t. $\forall \mathbf{p}^{(1)}, \boldsymbol{\alpha}^{(1)}, \mathbf{n}^{(1)}$ and $\mathbf{p}^{(2)}, \boldsymbol{\alpha}^{(2)}, \mathbf{n}^{(2)}$ satisfying

$$\left| p_i^{(1)} - p_i^{(2)} \right| < \varepsilon', \left| \alpha_{ji}^{(1)} - \alpha_{ji}^{(2)} \right| < \varepsilon', \left| n_i^{(1)} - n_i^{(2)} \right| < \varepsilon', \forall i, j,$$

we have $\left| U(\mathbf{p}^{(1)}, \boldsymbol{\alpha}^{(1)}, \mathbf{n}^{(1)}) - U(\mathbf{p}^{(2)}, \boldsymbol{\alpha}^{(2)}, \mathbf{n}^{(2)}) \right| < \varepsilon$.

*2)* For $\mathbf{p}(f), \{\alpha_{ji}(f)\}$ and $\{n_i(f)\}$, since bounded continuity implies uniform continuity, $\forall \varepsilon' > 0, \exists \delta > 0$, s.t. $\forall |f_1 - f_2| < \delta$, $f_1, f_2 \in [0,1]$, we have

$$\left| p_i(f_1) - p_i(f_2) \right| < \varepsilon', \left| n_i(f_1) - n_i(f_2) \right| < \varepsilon', \forall i,$$
$$\left| \alpha_{ji}(f_1) - \alpha_{ji}(f_2) \right| < \varepsilon', \forall i, j.$$

Now, combining 1) and 2), $\forall \varepsilon > 0$, divide $[0,1]$ into consecutive intervals $I_1, ..., I_M$ with lengths all less than $\delta$. $\forall m = 1, ..., M$, $\forall f \in I_m$, let

$$\overline{p}_i(f) = \min_{f \in I_m} p_i(f), \forall i,$$
$$\overline{\alpha}_{ji}(f) = \max_{f \in I_m} \alpha_{ji}(f), \forall i \neq j, \overline{n}_i(f) = \max_{f \in I_m} n_i(f), \forall i.$$

Thus, Property P1, P2 is satisfied, and $\forall f \in [0,1]$, $\forall i, j$,

$$\left| p_i(f) - \overline{p}_i(f) \right| < \varepsilon', \left| n_i(f) - \overline{n}_i(f) \right| < \varepsilon', \left| \alpha_{ji}(f) - \overline{\alpha}_{ji}(f) \right| < \varepsilon'$$
$$\Rightarrow \left| U(\overline{\mathbf{p}}(f), \overline{\boldsymbol{\alpha}}(f), \overline{\mathbf{n}}(f)) - U(\mathbf{p}(f), \boldsymbol{\alpha}(f), \mathbf{n}(f)) \right| < \varepsilon.$$

Thus, we have proved the lemma for bounded continuous power allocations and channel responses. To generalize it to piecewise continuous cases, simply use the fact that the number of discontinuities in $\mathbf{p}(f)$, $\{\alpha_{ji}(f)\}$ and $\{n_i(f)\}$ are *finite*. Thus, we can construct piecewise flat functions $\overline{\mathbf{p}}(f)$, $\{\overline{\alpha}_{ji}(f)\}$ and $\{\overline{n}_i(f)\}$ in every sub-interval with bounded continuity, and the values on the discontinuities do not have any impact on the power and rates, as they form a set of measure zero. ∎

## APPENDIX B

*Proof of Lemma 2:*

We want to show

$$\frac{df(x)}{dx} = \frac{2cx - (c^2 - x^2) \log(\frac{c+x}{c-x})}{x^2(c^2 - x^2)} \geq 0, \ x \in (0,1]$$

Since $c > 1$, it is equivalent to show

$$\frac{2cx}{c^2 - x^2} \geq \log(\frac{c+x}{c-x}), \ x \in (0,1].$$

14Let $g(x) = \dfrac{2cx}{c^2 - x^2}$ and $h(x) = \log(\dfrac{c+x}{c-x})$, $x \in [0, 1]$.

We have $\dfrac{dg(x)}{dx} = \dfrac{2c(c^2 + x^2)}{(c^2 - x^2)^2} = \dfrac{dh(x)}{dx} \dfrac{c^2 + x^2}{c^2 - x^2} \geq \dfrac{dh(x)}{dx}$.

Since $g(0) = h(0) = 0$, we have $g(x) \geq h(x)$, $x \in [0, 1]$.

Thus, $\dfrac{df(x)}{dx} \geq 0$, $x \in (0,1] \Rightarrow f(1) \geq f(x)$, $\forall x \in (0, 1]$. ∎

*Proof of Corollary 1.1:*

The flat FDMA power re-allocation for $K$ users is as follows (Figure 8): Each user $i$ re-allocates all its power within a disjoint sub-band $W_i' = p_i / \sum_{j=1}^{K} p_j$, with a flat PSD $p_i' = \sum_{j=1}^{K} p_j$.

With $\alpha_{ji} \geq 1/2$, $\forall j \neq i$, we have

$$R_1' = W_1' \log\left(1 + \frac{p_1'}{n_1}\right) = \frac{p_1}{\sum_{j=1}^{K} p_j} \log\left(1 + \sum_{j=1}^{K} p_j / n_1\right)$$

$$\geq \log\left(1 + \frac{p_1}{n_1 + \sum_{j=2}^{K} \frac{1}{2} p_j}\right) \geq \log\left(1 + \frac{p_1}{n_1 + \sum_{j=2}^{K} \alpha_{j1} p_j}\right) = R_1,$$

where the first inequality is by treating all the other users as one interferer, and using Theorem 1.

Similarly, $R_i' \geq R_i$, $\forall i = 2, ..., K$. ∎

APPENDIX C

*Proof of Lemma 4:*

With flat frequency sharing, the rates of user 1 and user 2 are

$$R_1 = \log(1 + \frac{p_1}{n + p_2 \alpha}), \quad R_2 = \log(1 + \frac{p_2}{n + p_1 \alpha}).$$

With a flat FDMA re-allocation, the rates of user 1 and user 2 become

$$R_1' = \frac{p_1}{p_1 + p_2} \log(1 + \frac{p_1 + p_2}{n}), R_2' = \frac{p_2}{p_1 + p_2} \log(1 + \frac{p_1 + p_2}{n}).$$

Straightforward calculations lead to

$$R_1 + R_2 \leq R_1' + R_2' \Leftrightarrow p_1 p_2 \left(\alpha^2(p_1 + p_2) - (2\alpha - 1)\right) \geq 0,$$

which implies the conclusion of Lemma 4. ∎

*Proof of Lemma 5:*

Clearly, the condition of $p$ implies $(p_1, p_2) \in \overline{P}_{FDMA}$.

First, we find the solution to the optimization problem with an equality sum-power constraint instead of inequality, i.e.,

$\max\ f(p_1, p_2)$, $s.t.\ p_1 + p_2 = p$, $p_1 \geq 0$, $p_2 \geq 0$.

With $p_1 + p_2 = p$,

$$f(p_1, p_2) = \left(\log(1 + \frac{p_1}{1 + \alpha(p - p_1)}) + \log(1 + \frac{p - p_1}{1 + \alpha p_1})\right) \triangleq \tilde{f}(p_1).$$

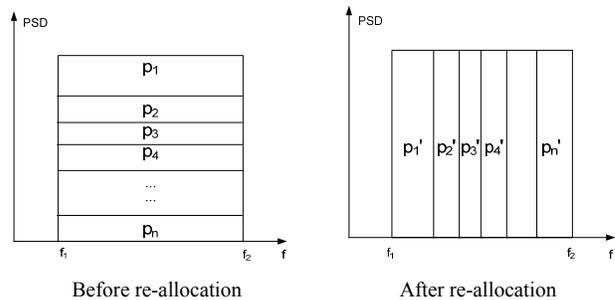

Fig. 8. PSDs before and after flat FDMA re-allocation of flat frequency sharing.

Straightforward calculations lead to

$$\frac{d}{dp_1}\tilde{f}(p_1) =$$

$$\frac{(1 + \alpha p)(\alpha^2 p + 2\alpha - 1)(p - 2p_1)}{(1 + \alpha(p - p_1))(1 + \alpha(p - p_1) + p_1)(1 + \alpha p_1)(1 + \alpha p_1 + p - p_1)},$$

Since $0 < p \leq 2\left(\dfrac{1}{2\alpha^2} - \dfrac{1}{\alpha}\right)$, when $0 \leq p_1 \leq p/2$, $\tilde{f}(p_1)$ is non-decreasing. Furthermore, note that $\tilde{f}(p_1) = \tilde{f}(p - p_1)$, i.e. $\tilde{f}(p_1)$ is *symmetric* about $p_1 = p/2$. Therefore, $\tilde{f}(p_1)$ takes the maximum value $2\log\left(1 + \dfrac{p/2}{1 + \alpha p/2}\right)$ when $p_1 = p/2$. With straightforward calculations, one can verify that $\log\left(1 + \dfrac{p/2}{1 + \alpha p/2}\right)$ is an *increasing concave* function of $p$.

Consequently, the constraint $p_1 + p_2 \leq p$ in the definition problem of $f^*(p)$ (5) can be equivalently replaced by $p_1 + p_2 = p$, and we have $f^*(p) = 2\log\left(1 + \dfrac{p/2}{1 + \alpha p/2}\right)$. ∎

*Proof of Theorem 6:*

1. $R^o \leq R^*$ (Converse).

It is immediately true, because the integrands in (17) and (18) by definition satisfies

$$R(\mathbf{p}(f), f) \leq R^*(\mathbf{p}(f), f), \forall f \in [0,1].$$

2. $R^* \leq R^o$ (Achievability).

Let $\mathbf{p}^*(f)$ be an optimal solution of (18) such that $\int_0^1 R^*(\mathbf{p}^*(f), f)df = R^*$. Then, $\forall \varepsilon > 0$:

By Lemma 2, based on $\mathbf{p}^*(f)$, $\{\alpha_{ji}(f)\}$ and $\{n_i(f)\}$, take a piecewise flat $\varepsilon$-approximation $\overline{\mathbf{p}}^*(f)$, $\{\overline{\alpha}_{ji}(f)\}$ and $\{\overline{n_i}(f)\}$, s.t.

$$\left|\int_0^1 \overline{R}^*(\overline{\mathbf{p}}^*(f), f)df - R^*\right| < \varepsilon,$$

where for every fixed $f \in [0,1]$, $\overline{R}^*(\mathbf{P}, f) \triangleq conv_{\mathbf{P}}\overline{R}(\mathbf{P}, f)$,

with $\overline{R}(\mathbf{P}, f) = \sum_{i=1}^{K} w_i \log \left( 1 + \dfrac{P_i}{\overline{n}_i(f) + \sum_{j \neq i} P_i \overline{\alpha}_{ji}(f)} \right)$ .

Based on the piecewise flat $\varepsilon$ - approximation, *in every flat sub-channel with a flat* $\vec{\mathbf{p}}^*(f)$, as in Theorem 5, $\overline{R}^*(\vec{\mathbf{p}}^*(f), f)$ can be achieved by further dividing this sub-channel into $K+1$ sub-bands, each applying flat power allocations. Denote the resulting allocation scheme by $\mathbf{p}^o(f)$ , achieving the same sum-rate

$$\int_0^1 \overline{R}(\mathbf{p}^o(f), f) df = \int_0^1 \overline{R}^*(\vec{\mathbf{p}}^*(f), f) df .$$

Then,

$$R^o \geq \int_0^1 R(\mathbf{p}^o(f), f) df$$
$$\geq \int_0^1 \overline{R}(\mathbf{p}^o(f), f) df = \int_0^1 \overline{R}^*(\vec{\mathbf{p}}^*(f), f) df > R^* - \varepsilon,$$

where the first inequality occurs because $\mathbf{p}^o(f)$ is a feasible solution of (17); the second inequality occurs because (by P2 from Lemma 1) $\overline{\alpha}_{ji}(f) \geq \alpha_{ji}(f), \forall i \neq j, n_i(f) \geq n_i(f), \forall i, \forall f$, i.e., the $\varepsilon$ - approximation worsens the channel responses, resulting in lower rates.

Finally, let $\varepsilon \to 0$ .                                                                     ∎

## REFERENCES

[1] Annapureddy, V.S.; Veeravalli, V.V., "Gaussian Interference Networks: Sum Capacity in the Low-Interference Regime and New Outer Bounds on the Capacity Region," *Information Theory, IEEE Transactions on*, vol.55, no.7, pp.3032-3050, July 2009.

[2] Bhaskaran, S.R.; Hanly, S.V.; Badruddin, N.; Evans, J.S.; , "Maximizing the Sum Rate in Symmetric Networks of Interfering Links," Information Theory, IEEE Transactions on , vol.56, no.9, pp.4471-4487, Sept. 2010.

[3] Boyd, S.P., Vandenberghe, L., "Convex Optimization," Cambridge University Press, 2004.

[4] Cendrillon, R.; Jianwei Huang; Mung Chiang; Moonen, M., "Autonomous Spectrum Balancing for Digital Subscriber Lines," Signal Processing, IEEE Transactions on , vol.55, no.8, pp.4241-4257, Aug. 2007.

[5] Cendrillon, R.; Wei Yu; Moonen, M.; Verlinden, J.; Bostoen, T., "Optimal multiuser spectrum balancing for digital subscriber lines," Communications, IEEE Transactions on , vol.54, no.5, pp. 922-933, May 2006.

[6] Mung Chiang, "Balancing transport and physical Layers in wireless multihop networks: jointly optimal congestion control and power control," Selected Areas in Communications, IEEE Journal on , vol.23, no.1, pp. 104-116, Jan. 2005.

[7] Mung Chiang, "Geometric programming for communication systems," Foundations and Trends in Communications and Information Theory, vol. 2, no. 1/2, pp. 1-156, August 2005.

[8] Ebrahimi, M.; Maddah-Ali, M.A.; Khandani, A.K.; , "Power Allocation and Asymptotic Achievable Sum-Rates in Single-Hop Wireless Networks," Information Sciences and Systems, 2006 40th Annual Conference on , vol., no., pp.498-503, 22-24 March 2006.

[9] Etkin, R.; Parekh, A.; Tse, D., "Spectrum sharing for unlicensed bands," Selected Areas in Communications, IEEE Journal on , vol.25, no.3, pp.517-528, April 2007.

[10] Etkin, R.H.; Tse, D.N.C.; Hua Wang, "Gaussian Interference Channel Capacity to Within One Bit," *Information Theory, IEEE Transactions on* , vol.54, no.12, pp.5534-5562, Dec. 2008.

[11] Gjendemsjo, A.; Gesbert, D.; Oien, G.E.; Kiani, S.G.; , "Optimal Power Allocation and Scheduling for Two-Cell Capacity Maximization," Modeling and Optimization in Mobile, Ad Hoc and Wireless Networks, 2006 4th International Symposium on , vol., no., pp. 1- 6, 03-06 April 2006

[12] Goldsmith, A., "Wireless Communications," Cambridge University Press, 2005.

[13] Hayashi, S.; Zhi-Quan Luo, "Spectrum Management for Interference-Limited Multiuser Communication Systems," Information Theory, IEEE Transactions on , vol.55, no.3, pp.1153-1175, March 2009.

[14] Zhi-Quan Luo; Shuzhong Zhang, "Dynamic Spectrum Management: Complexity and Duality," Selected Topics in Signal Processing, IEEE Journal of , vol.2, no.1, pp.57-73, Feb. 2008.

[15] Motahari, A.S.; Khandani, A.K.; , "Capacity Bounds for the Gaussian Interference Channel," *Information Theory, IEEE Transactions on* , vol.55, no.2, pp.620-643, Feb. 2009.

[16] Xiaohu Shang; Kramer, G.; Biao Chen; , "A New Outer Bound and the Noisy-Interference Sum–Rate Capacity for Gaussian Interference Channels," *Information Theory, IEEE Transactions on* , vol.55, no.2, pp.689-699, Feb. 2009

[17] Hongxia Shen; Hang Zhou; Berry, R.A.; Honig, M.L., "Optimal spectrum allocation in Gaussian interference networks," *Signals, Systems and Computers, 2008 42nd Asilomar Conference on* , vol., no., pp.2142-2146, 26-29 Oct. 2008.

[18] Tse, D.N.C.; Hanly, S.V., "Linear multiuser receivers: effective interference, effective bandwidth and user capacity," Information Theory, IEEE Transactions on , vol.45, no.2, pp.641-657, Mar 1999.

[19] Wei Yu; Cioffi, J.M., "FDMA capacity of Gaussian multiple-access channels with ISI," Communications, IEEE Transactions on , vol.50, no.1, pp.102-111, Jan 2002.

[20] Wei Yu; Ginis, G.; Cioffi, J.M., "Distributed multiuser power control for digital subscriber lines ," Selected Areas in Communications, IEEE Journal on , vol.20, no.5, pp.1105-1115, Jun 2002.

[21] Wei Yu; Lui, R., "Dual methods for nonconvex spectrum optimization of multicarrier systems," Communications, IEEE Transactions on , vol.54, no.7, pp. 1310-1322, July 2006.

[22] Yue Zhao; Pottie, G.J., "Optimal spectrum management in two-user symmetric interference channels," Information Theory and Applications Workshop, 2009 , vol., no., pp.256-263, 8-13 Feb. 2009.

[23] Yue Zhao; Pottie, G.J., "Optimal spectrum management in multiuser interference channels," Information Theory, 2009. ISIT 2009. IEEE International Symposium on , vol., no., pp.2266-2270, June 28 2009-July 3 2009.

[24] Yue Zhao; Pottie, G.J. , "Optimization of power and channel allocation using the deterministic channel model," *Information Theory and Applications Workshop (ITA), 2010* , vol., no., pp.1-8, Jan. 31 2010-Feb. 5 2010.

[25] Timothy M. Chan, "Optimal output-sensitive convex hull algorithms in two and three dimensions". Discrete and Computational Geometry, Vol. 16, pp.361–368. 1996.